\documentclass[a4paper]{aa}
\usepackage{aabib, epsfig, times}

\begin{document}

\title{The angular power spectrum of radio emission at 2.3 GHz}

\author{G.~Giardino\inst{1}, A.~J.~Banday\inst{2}, P.~Fosalba\inst{1},
  K.~M.~G\'orski\inst{3}, J.~L.~Jonas\inst{4}, W.~O'Mullane\inst{1}, 
  J.~Tauber\inst{1}}

\institute{Astrophysics Division -- Space Science Department of ESA, ESTEC,
  Postbus 299, NL-2200 AG Noordwijk, The Netherlands
\and
Max-Planck Institut f{\"u}r Astrophysik, Garching bei M\"{u}nchen,
                                             D-85741, Germany 
\and
ESO, Garching bei M\"{u}nchen, D-85748, Germany
\and
{Department of Physics \& Electronics, Rhodes University, PO Box 94,
  Grahamstown 6140, South Africa}
}

\offprints{G. Giardino} \mail{ggiardin@astro.estec.esa.nl}

\date{Received date ; accepted date}

\titlerunning{Angular power spectrum ...}
\authorrunning{G. Giardino et al.}

\abstract{We have analysed the Rhodes/HartRAO survey at
2326 MHz and derived the global angular power spectrum of Galactic
continuum emission. In order to measure the angular power spectrum of
the diffuse component, point sources were removed from the map by
median filtering.  A least-square fit to the angular power spectrum of
the entire survey with a power law spectrum $C_l \propto l^{-\alpha}$,
gives $\alpha = 2.43 \pm 0.01$ in the $l$ range $2-100$.  The angular
power spectrum of radio emission appears to steepen at high Galactic
latitudes and for observed regions with $|b| > 20^{\circ}$, the fitted
spectral index is $\alpha = 2.92 \pm 0.07$.  
We have extrapolated this result to 30 GHz (the lowest
frequency channel of Planck) and estimate that no significant
contribution to the sky temperature fluctuation is likely to come from
synchrotron at degree-angular scales.
\keywords{Radio continuum: ISM -- Surveys -- cosmic microwave
    background -- Techniques: image processing}} 
\maketitle

\section{Introduction}
\label{sec:intro}

The accuracy with which we are able to derive the cosmological
parameters from the angular power spectrum of the cosmic microwave
background (CMB) depends on the properties of Galactic and
extra-galactic foreground emission.  The future satellite missions
Planck (ESA) and MAP (NASA) designed to perform high precision
observations of the CMB have a wide frequency coverage in order to
disentangle the various components of the microwave sky and provide
clean CMB maps. The two satellites, therefore, will also provide new
high quality data on Galactic diffuse emission and extra-galactic
sources (\cite{Bersanelli96}). However before these observations
become available it is important to improve our knowledge of the
statistical properties of the foreground emission in order to quantify
the level of contamination of the CMB signal expected in the various
channels and at the different angular scales. This will allow more
realistic simulations of the missions to be performed and the analysis
techniques to be refined.

Below 50 GHz measurements of the CMB anisotropy are mostly affected by
Galactic radio emission: a combination of thermal bremsstrahlung from
ionized gas, synchrotron radiation from cosmic ray electrons and
possibly emission from spinning dust grains (\cite{Oliveira99}).  As
pointed out by \cite*{Tegmark00} when trying to recover the angular
power spectrum (APS) of the CMB, the uncertainties in the shape of the
APS of a foreground component is a greater source of error than the
uncertainties in the frequency dependence.  The only available
estimates of the APS of synchrotron emission comes from the analysis
of the 408-MHz survey of \cite*{haslam} and 1420-MHz survey of
\cite*{reich}.  They suggest a power-law spectrum with a spectral
index around 2.5$-$3.0 down to the maps' resolution limit $\sim
1^{\circ}$ (\cite{Tegmark96}; \cite{Bouchet96}; \cite{Bouchet99}).

Recently the Rhodes/HartRAO radio continuum survey at 2326 MHz
(\cite{Jonas98}) has become available. In this paper we use these new
data to derive the APS of the radio continuum at 2.3 GHz for all the
observed sky and for high Galactic latitude regions.  We then
extrapolate the high Galactic latitude results to predict the level of
contamination that might be expected from diffuse synchrotron emission
in the lowest frequency channel of Planck.

The paper is organized as follows.  In Sect.~\ref{sec:rhodes} the data
from the Rhodes/HartRAO survey are presented and the APS is derived for
different regions of the sky, after point source subtraction.  In
Sect.~\ref{sec:simulations} the simulations which were performed to
assess the reliability of our results are described.  Simulations and
results are discussed in Sect.~\ref{sec:disc}.

\section{The angular power spectrum of radio emission at 2.3 GHz}
\label{sec:rhodes}

The Rhodes/HartRAO radio continuum survey at 2326 MHz is a single-dish
survey of the sky with a resolution (at FWHM) of 20 arcmin
(\cite{Jonas98}). The survey made from the Hartebeesthoek Radio
Astronomy Observatory (HartRAO) in South Africa covers the declination
range $-83^{\circ} < \delta < 32^{\circ}$ for RA
$360^{\circ}-180^{\circ}$ and $-83^{\circ} < \delta < 13^{\circ}$ for
most of RA $180^{\circ}-0^{\circ}$ (in the RA range
$90^{\circ}-15^{\circ}$, $-80^{\circ} < \delta < 13^{\circ}$).  The
observations are spatially oversampled both in RA direction (6 arcmin)
and in declination direction (1.5 arcmin). The uncertainty in the
temperature scale is less than 5 per cent and the error in the absolute
zero level is less than 80 mK in any direction.
 
In order to analyse the global properties of the radio continuum
emission at 2326 MHz, the raw data of the Rhodes/HartRAO survey were
re-gridded into a HEALPix tessellation.  HEALPix is a Hierarchical,
Equal Area and iso-Latitude Pixelisation of the sphere designed to
support efficiently: local operations on the pixel set, a hierarchical
tree structure for multi-resolution applications and the global Fast
Spherical Harmonic transform (\cite{Gorski98}). This last property is
crucial for fast computation of the APS of a field defined on the
sphere.  The recent CMB maps obtained from the observations of the
balloon-born experiment Boomerang have also been pixelised according
to the HEALPix scheme (\cite{DeBernardis00}).

\begin{figure*}[!thbp]
  \begin{center}
    \leavevmode \psfig{file=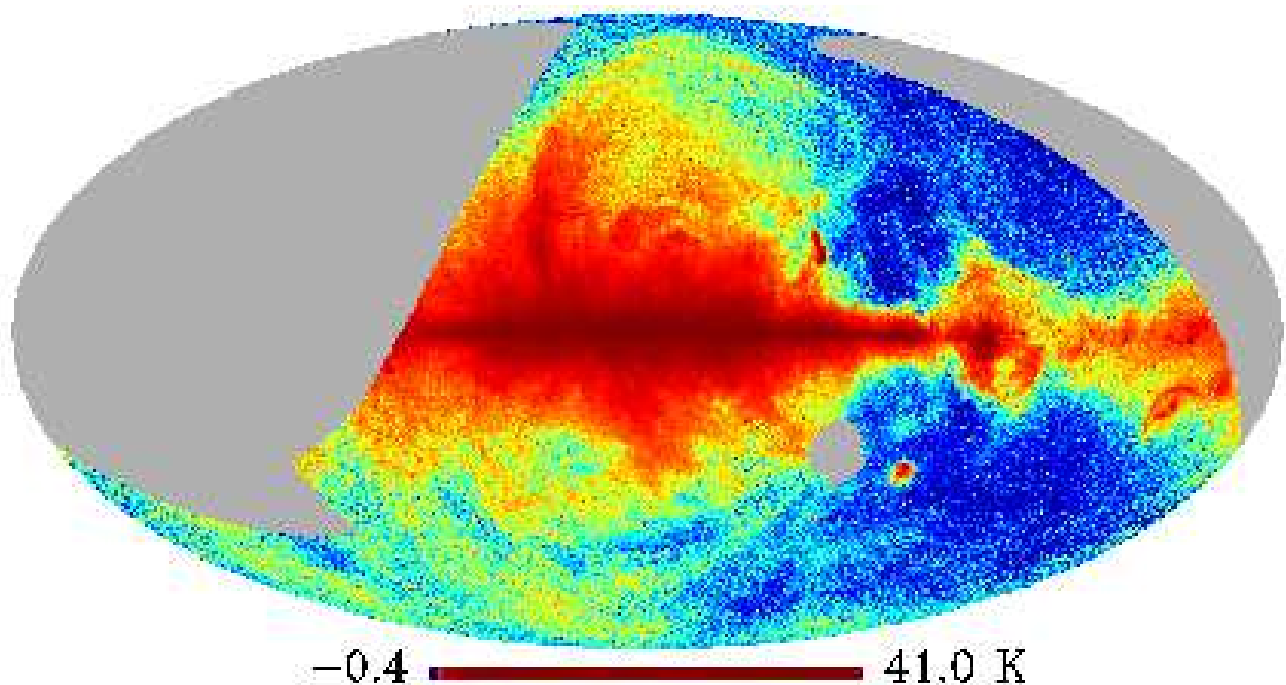,angle=0, width=\textwidth}
    \caption{Mollweide projection of the raw data from the Rhodes/HartRAO
    survey by Jonas et al.~(1998), re-sampled into an HEALPix map. The
    temperature scale has been histogram-equalised.}
    \label{fig:raw2326}
  \end{center}
\end{figure*}

In the HEALPix tessellation the angular size of the pixels is
determined by the {\em nside} parameter.  For the
Rhodes/HartRAO survey we used a HEALPix tessellation with {\em
  nside}=512 which corresponds to pixels of linear size 6.9 arcmin and
therefore provides an adequate re-sampling of the survey.  The
Mollweide projection of the raw data of the survey re-gridded into
HEALPix is displayed in Fig.~\ref{fig:raw2326}.  A great number of
point sources are clearly visible in the map and the survey appears to
be heavily affected by noise at the pixel scale.

\begin{figure*}[!thbp]
  \begin{center}
    \leavevmode \psfig{file=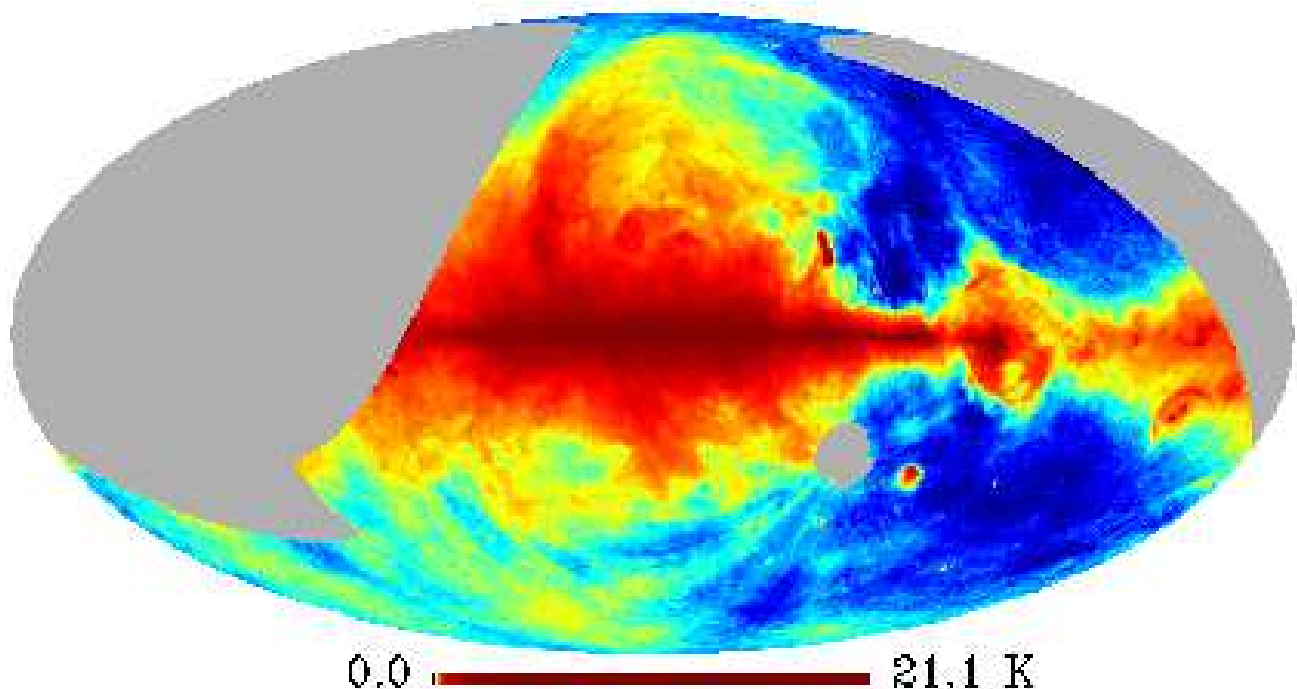,angle=0,width=\textwidth}
    \caption{Histogram-equalised Mollweide projection of the Rhodes/HartRAO
    survey, after median filtering}
    \label{fig:filt2326}
  \end{center}
\end{figure*}

The APS of a field on a rectangular-domain is given by the Fourier
components of the 2-point correlation function. The analogue in
spherical geometry are the $C_l$ components of the Legendre polynomial
expansion 
$$
C(\theta) = \frac{1}{4\pi} \sum_l (2l+1)C_lP_l(\cos \theta)
$$
If $T(\theta, \phi) = \sum_{lm} a_{lm}Y_{lm}(\theta, \phi) $
is the spherical harmonic decomposition of the field defined on the
sphere then the $C_l$ coefficients are simply given by:
\begin{equation}
C_l = \frac{1}{2l+1} \sum_{m=-l}^l |a_{lm}|^2
\label{eq:APS}   
\end{equation}
 
\begin{figure}[thbp]
  \begin{center}
    \leavevmode \epsfig{file=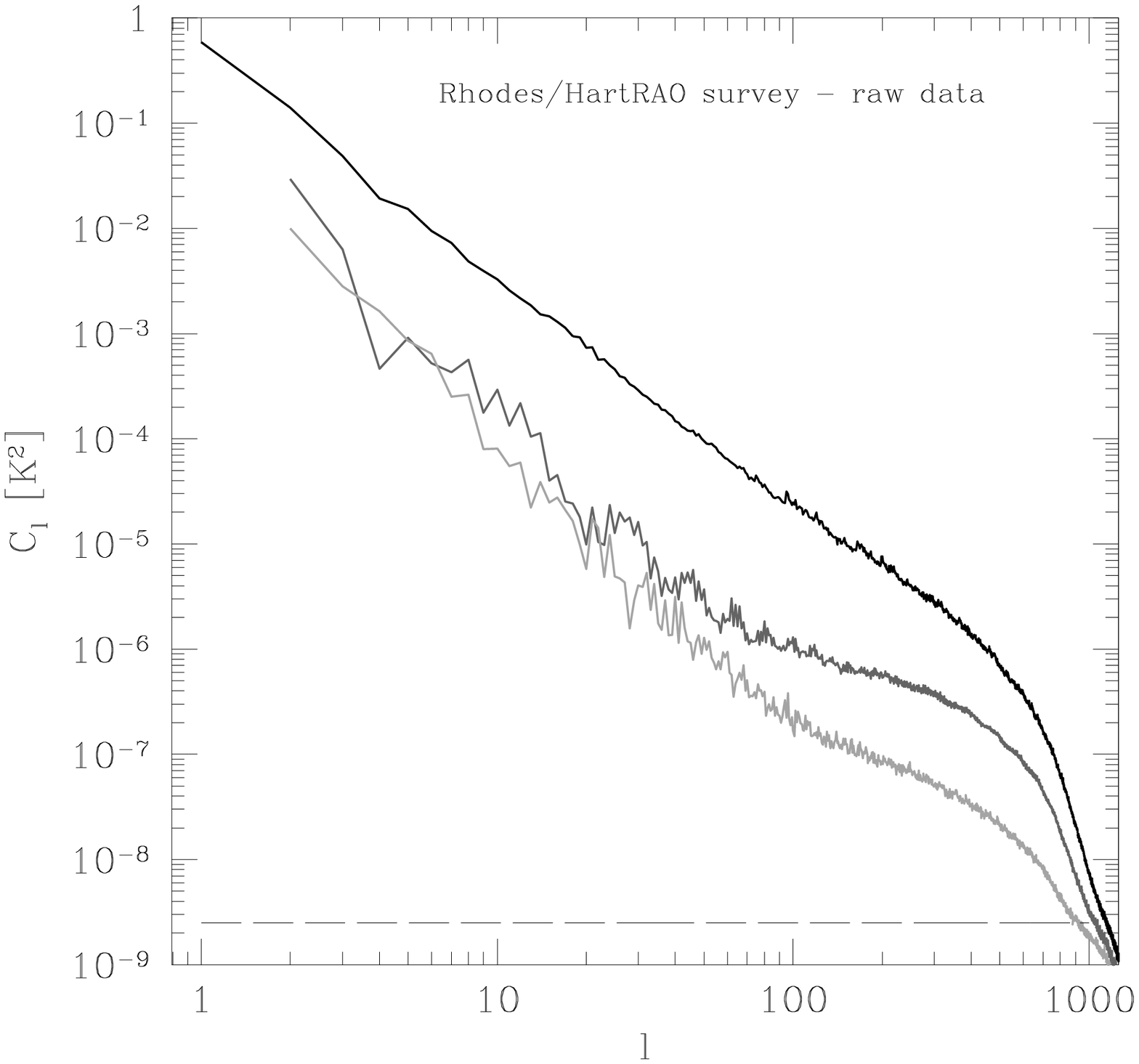, width=9.0cm, clip=}
    \caption{The angular power spectrum of the raw data from the
      Rhodes/HartRAO radio continuum survey at 2326 MHz for different
      Galactic cut-offs. Three spectra are shown: all observed sky
      (black line), galactic regions with $|b| > 10^{\circ}$
      (dark-grey line) and galactic regions with $|b| > 20^{\circ}$
      (light-grey line). The survey noise level is also shown (dashed line).}
    \label{fig:compRaw}
  \end{center}
\end{figure}

The APS of the Rhodes/HartRAO survey has been obtained by performing a
spherical harmonic analysis of the field in the HEALPix grid with the
ANAFAST program (the software is part of the HEALPix package available
at the web site {\tt http://www.eso.org/kgorski/healpix/}).  The
result of the analysis for the whole set of raw data are shown in
Fig.~\ref{fig:compRaw}, as well as the APS derived after imposing a
Galactic cut with ${\rm |b_{cut}|} = 10^{\circ}$ and the APS derived
for ${\rm |b_{cut}|} = 20^{\circ}$ \footnote{A Galactic cut with $b =
  {\rm b_{cut}}$, means that all the data in the latitude range ${\rm
    -b_{cut} < b < b_{cut}}$, have been set to zero}.
The presence of an equatorial cut across the sphere affects the
harmonic decomposition of the field.  Performing a cut in the space
domain is equivalent to multiplying the observed field by a window
function and, in the frequency domain, this is equivalent to the
convolution of the Spherical Transform of the field with the transform
of the window function. As a Galactic cut is a discontinuous window
function, its Spherical Transform is affected by aliasing. The
resulting APS is affected by slow oscillations, which reflect the
envelope of the transformed window function (which in this case is the
combination of the incomplete sky coverage with the Galactic cut), and
fast oscillations due to the aliasing signal. The oscillations are
clearly visible in Fig.~\ref{fig:compRaw}.

\begin{figure}[thbp]
  \begin{center}
    \leavevmode \epsfig{file=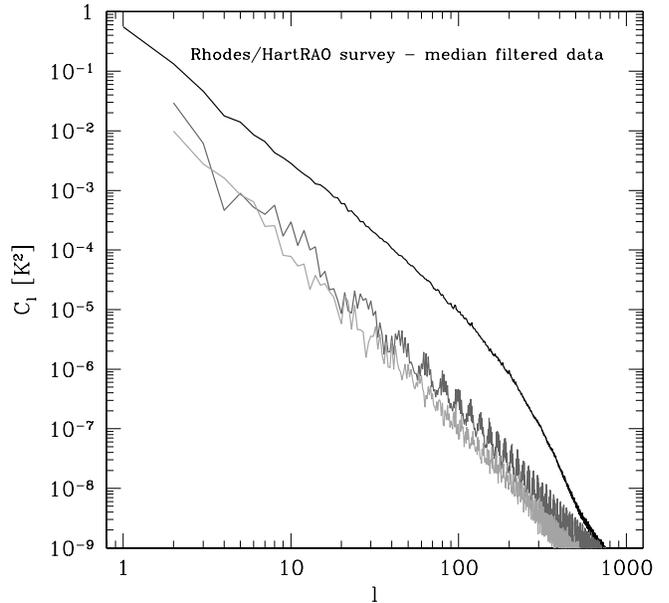, width=9.0cm, clip=}
    \caption{The angular power spectrum 
      of the data from the Rhodes/HartRAO survey at 2326 MHz, after the
      point sources have been removed using a median filter. Line shading
      as in previous figure.}
    \label{fig:compFilt}
  \end{center}
\end{figure}

The cylindrical symmetry of a Galactic cut manifests itself as a
``coherent'' set of terms in m=0, therefore the effect of this window
function on the measured APS can be mitigated by implementing a new
definition of the APS in which one sets the $a_{l0}$ terms to zero,
that is:
\begin{equation}
C_l = \frac{1}{2l} \left(\sum_{m = -l}^{m=-1} |a_{lm}|^2 
+ \sum_{m = 1}^{m=l} |a_{lm}|^2\right)~~~l\neq 0
\label{eq:ModAPS}   
\end{equation}
Eq.~\ref{eq:ModAPS} is the definition (which we shall refer to as
Modified APS, hereafter) that has been used to obtain the power spectra
from the data with a Galactic cut in Fig.~\ref{fig:compRaw} and
Fig.~\ref{fig:compFilt} (grey-lines).

The window function of a Gaussian beam with dispersion $\sigma_b$ is given by:
$$
W_l \approx \exp[-\frac{1}{2}\sigma_b^2l(l+1)]
$$ which means that for a beam with a FWHM of 20 arcmin the beam
suppression of the sky APS starts becoming appreciable (transmission
factor $<0.9$) at $l
\sim 200$ ($\sigma_b= {\rm FWHM}/\sqrt{8\ln2}$). 
We model the APS of this component of the radio sky as a single power
law:
$$
C_l = A^2l^{-\alpha}
$$ Up to $l=200$, a linear least-square fit to the APS of all raw data
(black line in Fig.~\ref{fig:compRaw}) gives $A = 0.63 \pm 0.14$ K and
$\alpha = 2.11 \pm 0.06$.  The APS at high Galactic latitude cannot
clearly be described by a single power law up to $l=200$, however if
one attempts a linear fit up to $l = 100$, where the curves flatten,
one obtains: $A = 0.2\pm 0.1$ K and $\alpha = 2.38\pm 0.05$, for ${\rm
  |b|>10^{\circ}}$ (dark-grey line), and $A = 0.2\pm 0.1$ K and $\alpha
= 2.71\pm0.03$ for regions with ${\rm |b| > 20^{\circ}}$ (light-grey
line). The noise level of the survey data, as given by \cite*{Jonas98},
is also shown in Fig.~\ref{fig:compRaw} (dashed line). From the diagram
it can be seen that the noise level is well below the signal, in the
$l$-range considered.

The flattening of the spectrum at intermediate values of $l$ is due to
the presence of point sources in the map.  In fact, the APS of a
population of point sources randomly distributed in the sky is
Poissonian, that is a spectrum with constant power at all $l$. As the
fluctuation amplitude of the diffuse emission diminishes at small
angular scales, the Poissonian component due to the point sources
becomes dominant.

Therefore, in order to analyse the angular power spectrum of the
Galactic diffuse emission the point sources should be removed from the
maps.  Each point source in the map has a profile reflecting the
point-spread function (PSF) of the observing beam. On a sufficiently
resolved map one could attempt to remove the sources by fitting the
pixel temperatures with the PSF. However at this low resolution several
sources may be present in any given pixel, thus precluding this treatment.
As a way of suppressing the point sources from the data we have made
use of a median filter, that is, a convolution of the map with a median
box of 9$\times$9 pixels.  A box of size 9 pixels corresponds to the
telescope beam size just over 3$\sigma_b$.  Median filtering, is an
effective and easy-to-implement method to remove point sources and
suppress random instrumental noise (e.g. \cite{Pratt91}), without
losing information at scales greater than three times the beam FWHM.
The image of the Rhodes/HartRAO survey after being filtered is
displayed in Fig.~\ref{fig:filt2326}.

The angular power spectra of the Rhodes/HartRAO survey for different
latitude cut-off, after applying median filtering, are plotted in
Fig.~\ref{fig:compFilt}.  For all the observed data, up to $l = 100$,
the best-fit power-law to the APS has a amplitude $A =
0.9 \pm 0.2$~K and spectral index $\alpha = 2.43 \pm 0.01$.  For CMB
studies one is particularly interested in the properties of radio
emission at high Galactic latitudes because this is where the Galactic
signal is weaker and where sensitive CMB measurements can be made.  At
this frequency the high Galactic latitude diffuse emission is dominated
by synchrotron radiation. Indeed the power spectrum of the emission
appears to steepen if we mask out the Galactic plane. For a Galactic
cut with $|b_{\rm cut}| = 10^{\circ}$, a linear least-square fit to
the APS up to $l =100$ (dark-grey line) gives $A=0.3\pm0.2$~K and
$\alpha = 2.81\pm 0.06$; for $|b_{cut}| = 20^{\circ}$ (light-grey line)
one obtains $A=0.3\pm0.2$~K and $\alpha = 2.92\pm 0.07$. The quoted
error takes into account an assessment of the uncertainties introduced
by the partial sky coverage, the presence of a Galactic cut and median
filtering. We have evaluated this error by performing various
simulations, as described in Sect.~\ref{sec:simulations}. Note from
Fig.~\ref{fig:compFilt} how the rapidly falling APS at high Galactic
latitude becomes dominated by the aliasing noise from the equatorial
cut off at $l > 200$: suppression of the sky signal by the beam-window
function is no longer visible.

\subsection{Point-source contribution}
\label{sec:pscontrib}

The difference map of the raw data and the median-filtered data gives
the map of the point sources plus the high-spatial-frequency
instrumental noise. We have computed the APS of this difference map at
high Galactic latitude and compared it with the APS from the source
counts expected at 2.3 GHz.  

\cite*{Toffolatti98} report the point-source differential counts at
1.4, 5 and 8.44 GHz, for flux ranges between $10^{-5}$ and 10 Jy. It
can be seen from Fig.1 of \cite*{Toffolatti98} that a differential
source count distribution of the form
\begin{equation}
dN/dS = {\rm 150S^{-2.5} ~sr^{-1} Jy^{-1}}
\label{eq:sc}   
\end{equation}
can be taken as a reasonable approximation of the source counts at 2.3
GHz in the flux range $0.1-10$~Jy.  Sources with lower flux levels will
not contribute significantly to the fluctuation of antenna temperature
in a 20 arcmin beam. The reason being that there are so many of these
sources in a 20 arcmin beam that changing pointing does not cause
significant changes to the average signal (\cite{ftdz}).  In addition,
the estimated receiver noise level for the Rhodes/HartRAO survey is 25
mK and the measured Point Source Sensitivity (PSS) for the survey is
9.8 Jy/K. This means that point sources with fluxes lower than 0.2 Jy
at 2.3 GHz are below the map noise level.

We produced 10 realisations of 40,000 point sources with source counts
of the form given in Eq.~\ref{eq:sc}, in the flux range $0.1-100$ Jy,
randomly distributed in the sky\footnote{In order to assess whether 10
  realisations were enough to characterize the population of point
  sources we generated another 20 realisations and found that the mean of
  all the 30 realisations was within one standard deviation of the 3
  subsamples of 10 realisations each}.  We convolved the realisations
with a Gaussian beam with 20 arcmin FWHM\footnote{The convolution was
  performed in the spatial-domain, to prevent any aliasing effect due
  to the point-like nature of the source signal.  The convolving
  Gaussian beam was truncated at $5\sigma$} and computed the APS for
each one.  The APS of a randomly distributed population of point
sources is given by
\begin{equation}
C_l = \int_0^{S_c} \frac{dN}{dS}S^{2}dS
\label{eq:scint}   
\end{equation}
where ${S_c}$ is the flux cut (\cite{Tegmark96}). Therefore the
amplitude of the APS of different realisations of point-sources with the
same differential source counts distribution will vary slightly
accordingly to the fluxes of the brightest sources present.

\begin{figure}[thbp]
  \begin{center}
    \leavevmode \epsfig{file=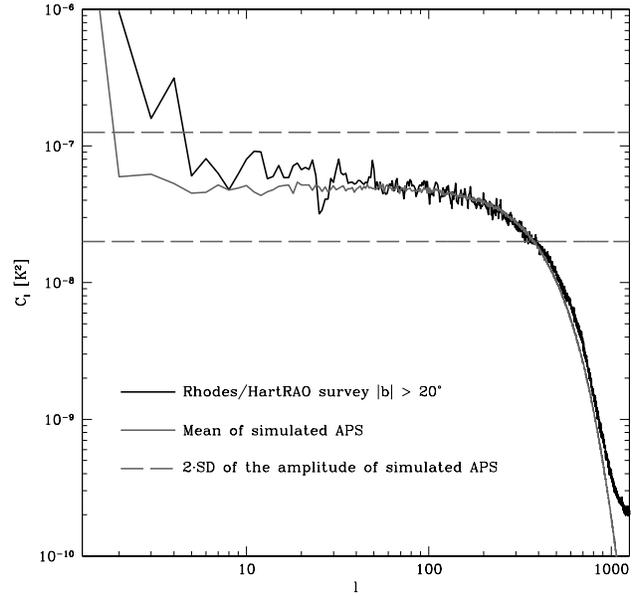,
    width=9.0cm, clip=}
    \caption{APS of the point sources extracted from the Rhodes/HartRAO
      survey by subtracting the median filtered data from the raw data
      (black solid-line), away from the galactic plane. The mean of the
      APS of 10 realisations of a random population of point sources
      with differential source counts of the form given in Eq.~3 is
      also reported (grey solid-line). The grey dashed lines give
      $2\cdot$SD (standard deviation) of the amplitude of the APS from
      the simulations. Model and observations are consistent with each
      other.}
    \label{fig:PSlevel}
  \end{center}
\end{figure}

In Fig.~\ref{fig:PSlevel} we show the APS of the sources extracted from
the Rhodes/HartRAO survey by median filtering for sky regions with $|b|
> 20^{\circ}$ (black line).  The results can be compared with the mean
APS of the 10 realisations of sources with differential counts
distribution given by Eq.~\ref{eq:sc} (continuous grey line).  In the
figure the standard deviation (SD) of the amplitude of the APS of the
10 realisations of point sources is also reported (dashed grey line).
The APS of the point sources extracted from the Rhodes/HartRAO survey
is consistent with a population of point sources having differential
source counts of the form given in \ref{eq:sc}, randomly distributed in
the sky.

Comparing angular power spectra is not the ideal way of comparing
source count distributions. As can be seen from Eq.~\ref{eq:scint}
the $C_l$ coefficients are given by the integral of the distribution up
to a flux-cutoff, therefore different source count distributions can
give angular power spectra of the same amplitude.  However the APS of
the point-source distribution is the quantity used to predict the level
of contamination to CMB anisotropy coming from extra-galactic point
sources (\cite{Toffolatti98}; \cite{Hobson99}). This result indicates
that the basic assumptions on the spatial and flux distribution of
extra-galactic radio sources used in CMB studies reproduce correctly the
observations at 2.3 GHz made with a 20-arcmin FWHM beam.

\section{Simulations} 
\label{sec:simulations}

The incomplete sky coverage of the Rhodes/HartRAO data, point source
suppression by median filtering and the presence of a Galactic cut
affect the accuracy with which the angular power spectrum at high
Galactic latitude may be recovered.  In order to assess the impact of
these factors on the estimates of the APS of diffuse radio
emission we have performed a series of simulations and reproduced the
processing procedure used on the real data.  

We have produced two sets of simulations: one for an input APS with a
spectral index of $\alpha= 2 $ and the other for an input APS with
$\alpha= 3$. For each
set we generated 10 different realisations of a Gaussian random field
on the sphere having that input APS. For this purpose we used the
SYNFAST software (also part of the HEALPix software package). All the
realisations were generated at a resolution of 20 arcmin (FWHM) and
sampled into HEALPix maps with {\em nside} = 512, as in the case of
the Rhodes/HartRAO map. From these we computed the APS for three
different conditions: total sky coverage, Galactic plane cutoff with
$|b_{\rm cut}| = 10^{\circ}$ and partial sky coverage plus Galactic
cutoff ($|b_{\rm cut}| = 10^{\circ}$). For an input APS with $\alpha=
3$ we also considered the case in which a population of spatially
random point sources is added onto the simulated diffuse emission.

\subsection{Diffuse emission only}
\label{sec:diffonly}

\begin{figure*}[thbp]
  \begin{center}
    \leavevmode \epsfig{file=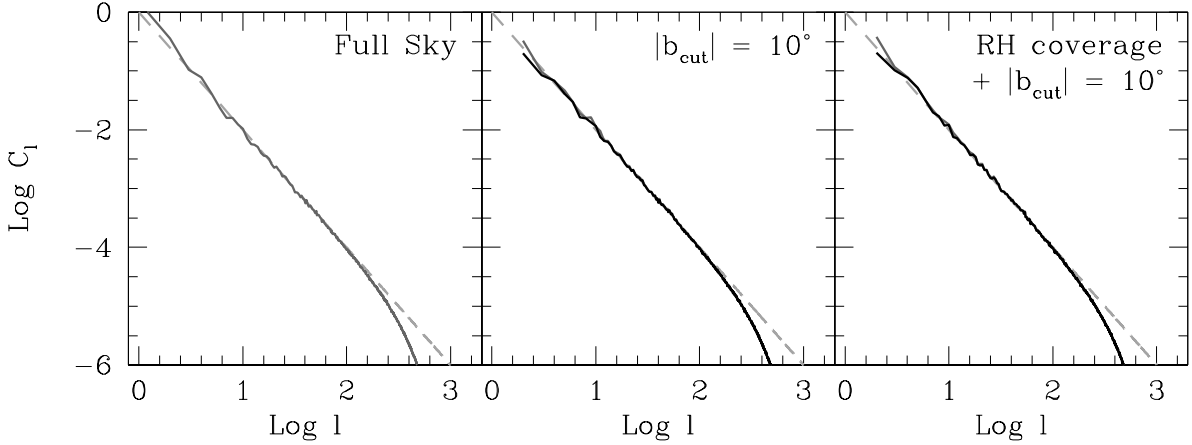, width=16.0cm, clip=}
    \caption{Mean recovery of the APS from 10 different realisations of an
    input APS with spectral index $2$ (dashed light-grey line), for
    the three different cases considered: full sky coverage, galactic
    cut at  $|b| = 10^{\circ}$ and Rhodes/HartRAO sky coverage plus
    galactic cut at  $|b| = 10^{\circ}$. The dark-grey lines refer to
    the standard definition for the APS, the black lines to the
    Modified APS.}
    \label{fig:MeanCll-2}
  \end{center}
\end{figure*}

In Fig.~\ref{fig:MeanCll-2} the means of the derived APS from 10
different realisations of a spectrum with spectral index $\alpha = 2$
are shown. The curves are for the three different cases considered,
that is deriving the APS from full sky maps of the realisations, from
maps with a Galactic cutoff with $|b_{\rm cut}| =10^{\circ}$ and from
maps with coverage identical to that of the Rhodes/HartRAO survey plus
a Galactic cutoff. With a Galactic cutoff both the usual definition
for the APS and the Modified APS definition have been used.  As seen
from Fig.~\ref{fig:MeanCll-2}, in this case, the incomplete sky
coverage and the presence of a Galactic cutoff do not affect the
recovery of the input APS in a substantial way, and there is no
appreciable difference between the standard APS definition and the
Modified definition.

\begin{figure*}[thbp]
  \begin{center} \leavevmode \epsfig{file=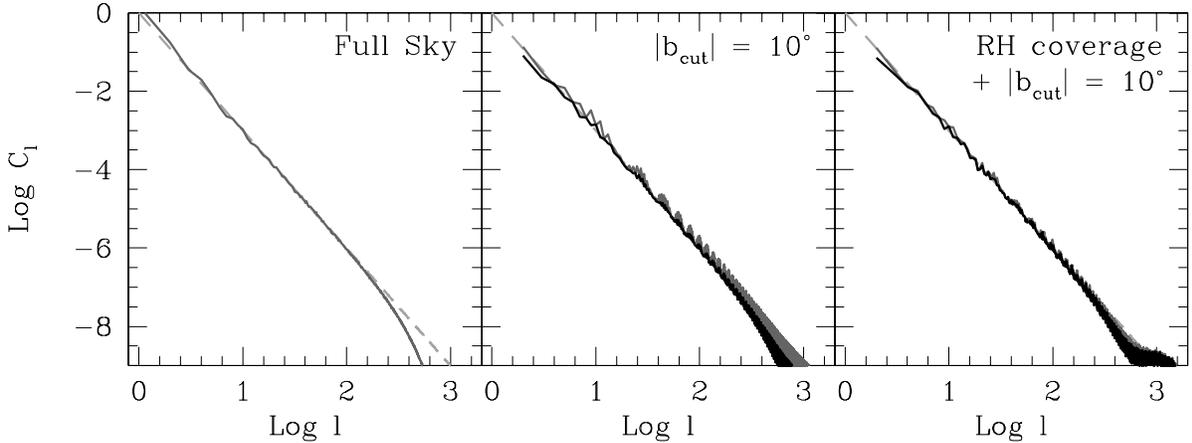,
    width=16.0cm, clip=} \caption{The mean of the spectra recovered
    from 10 different realisations of an input APS with spectral index
    $-3$. The curves are as in Fig.~6.}
    \label{fig:MeanCll-3} \end{center}
\end{figure*}

In order to quantify the accuracy of the recovery we have computed the
best-fit power law of the spectra obtained from the 10 realisations,
in the three different cases of sky coverage. The amplitude and
spectral index of these best-fit power laws can be compared with the
true amplitude and spectral index of the APS used as input for the 10
realisations. The standard deviation of the derived amplitudes and
spectral indices with respect to the true (input) values provides an
estimate of the standard error on the recovered amplitude and spectral
index.  The results are summarised in Table~\ref{tab:recovCl}. For a
full sky coverage the uncertainty in the measured spectral index over
the $l$-range 2--200 is of about 2\% as can be seen simply by
taking the standard deviation over the mean. An incomplete sky
coverage such as the one of the Rhodes/HartRAO survey plus the
introduction of a Galactic latitude cutoff brings this uncertainty up
to 4\%. The error on the amplitude of the power law is not
significantly affected by a diminished sky coverage.

\begin{table}
\begin{center}
  \footnotesize
\begin{tabular}{|c|c c c c c|}
\hline
~ & ~ & $A$ & ~ & $\alpha$ & ~\\
\hline
input $\alpha$ & Sky coverage &  mean & SD &  mean & SD \\
\hline
~ & Full Sky & 1.0 & 0.5 & 2.04 & 0.05 \\
2 & ${\rm |b_{cut}| = 10^{\circ}}$ & 1.0 & 0.5 &2.04 & 0.06 \\
~ & RH  $+{\rm |b_{cut}| = 10^{\circ}}$ & 1.0 & 0.5 & 2.07 & 0.08\\
\hline
~ & Full Sky & 1.1 & 0.7 & 3.04 & 0.05 \\
3 & ${\rm |b_{cut}| = 10^{\circ}}$ & 1.1 & 0.7 &3.02 & 0.04 \\
~ & RH $+{\rm |b_{cut}| = 10^{\circ}}$ & 1.1 & 0.8 & 3.05 & 0.06\\
\hline
\end{tabular}
\end{center}

\caption{Mean and standard deviation of the recovered amplitude and
spectral index from 10 different sky realisations of an input APS with
amplitude $A=1.0$~K and spectral index $\alpha = 2$ (top panel) and
$\alpha = 3$ (lower panel). Three different cases have been
considered: full sky coverage, Galactic cut at $|b_{\rm cut}| =
10^{\circ}$ and Rhodes/HartRAO sky coverage plus Galactic cut.  $l \in
[2, 200]$.}
\label{tab:recovCl}
\end{table}

The same cases of sky coverage have been considered also for an input
APS with spectral index $\alpha =3$. Fig.~\ref{fig:MeanCll-3} shows the
mean of the recovered spectra from 10 different sky realisations of the
input APS. In this case the derived spectra are more strongly affected
by a Galactic cutoff and an incomplete sky coverage.  The modulation of
the APS caused by introducing the Galactic plane cutoff is clearly
visible in the figure (middle panel). The figure also shows how the use
of the Modified APS definition helps in reducing such aliasing. 

The convolution of the field Spherical Transform by the Spherical
Transform of the sky coverage window (which is undersampled) introduces
a coupling between the $a_{lm}$ coefficients of the field, which is
stronger for fields with steeper APS, simply because the gradient in
$l$ is stronger. This is the reason why an input APS with $\alpha =2$ is
barely affected by the incomplete sky coverage, while this has such a
strong impact on the measure of an input APS with $\alpha =3$.  In
practice the coupling introduced by the sky-coverage window means that
no APS with spectral index steeper than $-3.5$ can be recovered in the
presence an equatorial cutoff or large areas of missing data.

However, despite the aliasing noise introduced by the Galactic plane
cutoff and the incomplete sky coverage, one can still recover the
spectral index of an input APS with $\alpha \sim 3$. The
parameters resulting from least-square fits to the Modified APS
obtained from the 10 realisations are reported also in
Table~\ref{tab:recovCl}.  The percentage error on the recovery of the
input spectral index from a full-sky observation, is in this case
about 1.5\% and does not increase substantially with decreasing sky
coverage.

\subsection{Diffuse emission and point sources}
\label{sec:diff+psources}

In Sect~\ref{sec:pscontrib} we have described how we generated 10
realisations of a spatially random population of point sources with
differential source counts of the form given in Eq.~\ref{eq:sc}. From
\cite*{Toffolatti98} one can see how this expression of the
differential source counts approximate quite well the extra-galactic
source counts that are expected at 2.3 GHz, in the flux range $0.1 -
10$ Jy. Sources with flux lower than 0.1 Jy do not contribute
significantly to the temperature
fluctuations in the Rhodes/HartRAO survey.  The 10 different
realisations of point sources were then added onto 10 different
realisations of an input APS with a spectral index of $\alpha = 3$,
normalised to the amplitude of the APS of the Rhodes/HartRAO survey
measured for a Galactic cutoff with ${\rm |b_{cut}| = 10^{\circ}}$.
Each resulting map, which mimics the superposition of diffuse emission
with a Poissonian population of point sources was then convolved with a
median filter of size 9$\times$9. The APS and the Modified APS of the
filtered map was then computed. This APS represents a recovery of the
spectrum of the diffuse component and can be directly compared with the
input spectrum of the diffuse component to assess the systematic error
introduced by median filtering the map.

\begin{figure}[thbp]
  \begin{center} \leavevmode \epsfig{file=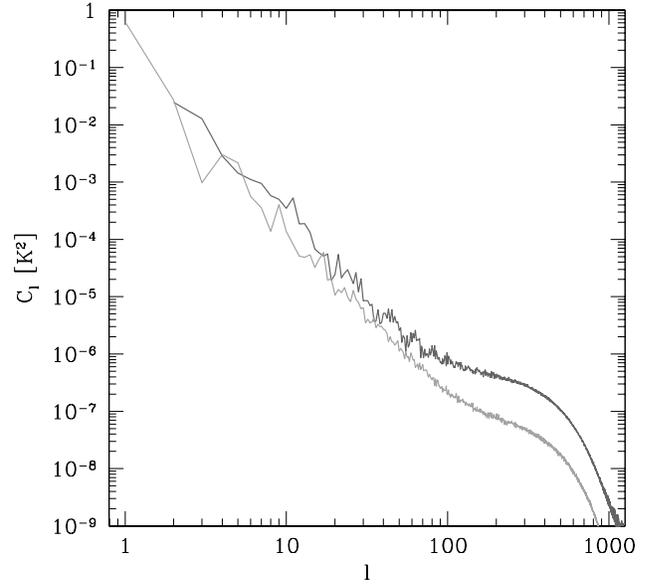,
    width=9.0cm, clip=} 
	\caption{APS of 2 simulated skies containing a
    diffuse component with power law spectrum with $\alpha =3$ and a
    population of point source of the form given in Eq.~3. Note the
    similarity with the APS of the raw data at high Galactic latitude
    shown in Fig.~3. In these two particular cases the spectral index
    up to $l=100$ are $\alpha = 2.49\pm0.02$ and $\alpha =
    2.79\pm0.02$ respectively, after median filtering the maps one
    obtains $\alpha =3.04 \pm 0.07$ and $\alpha = 2.96 \pm 0.07$.}
    	\label{fig:compWithSim} 
 \end{center}
\end{figure}

In Fig.~\ref{fig:compWithSim} the angular power spectra of two simulated
skies are shown on the same scale as the data in
Fig.~\ref{fig:compRaw}. As in the case of the real data the
flattening of the spectrum at high $l$ due to the Poissonian APS of the
point source component is clearly recognizable.  These 2 particular
cases of simulated skies mimic quite closely the spectra of the raw
data in Fig.~\ref{fig:compRaw}.  In these two cases the
fitted-spectral index up to $l=100$ of the input maps (diffuse
emission plus sources) are respectively $\alpha = 2.49 \pm 0.02$ and
$\alpha = 2.79\pm 0.02$\footnote{the fit error is given in this case},
and after median filtering one obtains $\alpha = 3.04 \pm 0.07$ and
$\alpha = 2.96 \pm 0.07$, which are good recoveries of the true spectral
index of the simulated diffuse emission.

\begin{figure*}[!thbp]
  \begin{center}
    \leavevmode \epsfig{file=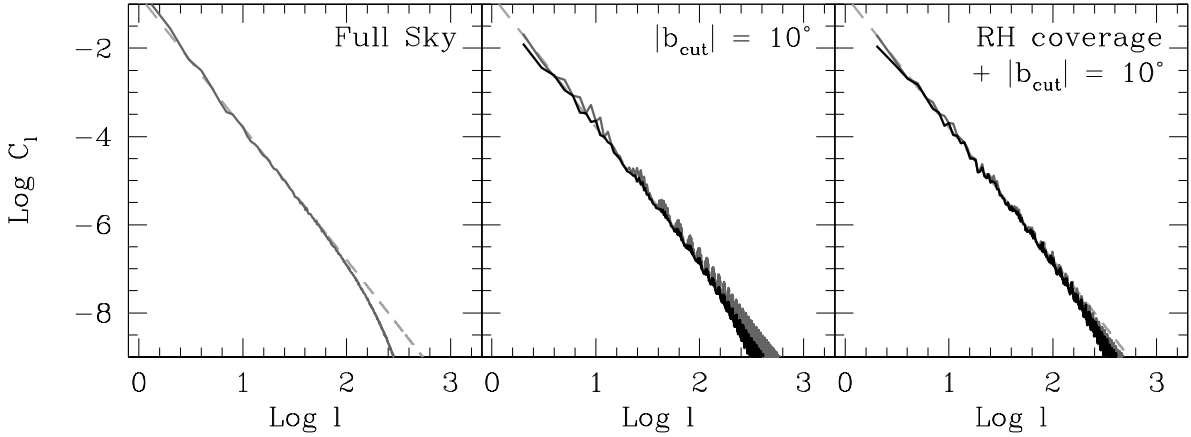, width=16.0cm, clip=}
    \caption{Mean of the spectra from 10 different realisations of an
      input APS with spectral index $\alpha=3$ (dashed light-grey line) plus a
      population of point sources with differential source counts of
      the form given in Eq.~3. The maps were first processed
      by median filter to remove the point source components. The three
      different cases of sky coverage considered are shown. The
      dark-grey lines refer to the standard definition for the APS, the
      black lines to the Modified APS.}
    \label{fig:MeanCll-3MF}
  \end{center}
\end{figure*}

Fig.~\ref{fig:MeanCll-3MF} shows the Mean APS of the diffuse component
recovered by median filtering in the three different cases of sky
coverage considered. The median filtering suppression of the high
frequency signal starts to affect the APS at $l\sim 100$. Up to this
multipole order the filtering process allows the correct spectral
index of the input APS to be recovered, with an error lower than 3\%,
as taken from Table~\ref{tab:Cll-3MF}. The average spectral index of
the spectra of the 10 maps before median filtering is 2.8, with a
standard deviation of $0.1$ for $l \in [2, 100]$.

\begin{table}
\begin{center}
\footnotesize
\begin{tabular}{c c c c c c c}
\hline
~ & $A$ & ~ & $\alpha$ & ~ \\
\hline
Sky coverage &  mean & SD &  mean & SD \\
\hline
Full Sky & 0.3 & 0.2 & 3.04 & 0.08 \\
${\rm |b_{cut}| = 10^{\circ}}$ & 0.3 & 0.2 &3.00 & 0.07 \\
RH $+{\rm |b_{cut}| = 10^{\circ}}$ & 0.3 & 0.2 & 3.02 & 0.07\\
\hline
\end{tabular}
\end{center}

\caption{Mean and standard deviation of the recovered amplitude and
spectral index of the APS of the diffuse component from 10 different
sky realisations. Each realisation consisted of a realisation of an
input APS with amplitude $A=0.3$~K and spectral index $\alpha=3$ plus
a realisation of a spatially random population of point sources with
source counts of the form given in Eq.~3. The mean spectral index of
the APS of the full-sky unprocessed maps was $2.8 \pm 0.1$. $l\in
[2, 100]$.}
\label{tab:Cll-3MF}
\end{table}
 
\section{Discussion}
\label{sec:disc}

The analysis of the Rhodes/HartRAO data shows that the global APS of
synchrotron emission at high Galactic latitudes ($|b| >20^{\circ}$) is well
approximated by a power law spectrum having spectral index $\alpha =
2.92 \pm 0.07$. This result at 2.3 GHz is consistent with the results
of \cite*{Tegmark96} and \cite*{Bouchet96} who found a spectral index
$\alpha$ of $2.5-3$ up to $l\sim 100$ from the analysis of square
patches of the 408-MHz survey. \cite*{Bouchet99} also found $\alpha
\sim 3$ for $l\geq 100$ from the analysis of patches of the 1420-MHz
survey.  

\cite*{Lasenby97} found $\alpha \sim 2$ by analysing the 408-MHz and
the 1420-MHz surveys in the ``Tenerife'' patch of sky.  This is a
$60\times 10$ degree patch of the northern hemisphere at high Galactic
latitude.  This could be a local property of that region of the sky.
Alternatively, it could be due to the effect of point sources: Lasenby
did not attempt to subtract point sources (private communication) and
this may have the effect of flattening the spectrum.  From
Fig.~\ref{fig:compRaw} one can see that without point source
subtraction the APS of the synchrotron emission has a slope of
approximately $-2$ in the $l$-range 50--150, which is the $l$-range
that can be probed with a patch of sky of the size of the ``Tenerife
patch'', at a resolution of $\sim 1^{\circ}$.

A spectral index of $\alpha \sim 3$ has also been derived for the APS of
Galactic dust emission (\cite{Gautier92};
\cite{Tegmark96}; \cite{Wright98}) and  this is the same slope measured
for the APS of single velocity neutral hydrogen in the Galaxy
(\cite{Crovisier83}). However, \cite*{Schlegel98} report a flatter
slope of $\alpha =2.5$ for dust emission from the combined DIRBE and
IRAS dust maps, at high Galactic latitudes.

The simulations we have performed showed that incomplete sky coverage
and a Galactic cut do not introduce significant errors when deriving
the spectrum of an input field with $C_l \propto l^{-2}$. This is
reassuring since $\alpha \sim 2$ is approximately the spectral index of
the APS of the CMB.  On the other hand the incomplete sky coverage of
the Rhodes/HartRAO survey and a Galactic cut seriously degrade the
quality of the spectrum that can be derived from an input field with a
spectral index steeper than 3.  Therefore, while spherical harmonic
decomposition is a suitable tool to analyse the statistical properties
of CMB emission with incomplete sky coverage, it is not ideal when one
tries to analyse fields whose spectra have index $\alpha \ga 3$.  This
highlights the need to refine the analysis tools in view of the new
observations of the Galactic diffuse emission that will be carried out
by Planck and MAP satellites.

Indeed, one limit of our simulations is that the synthesized diffuse
emission is a realisation of a random Gaussian field with a given APS
while the Galactic signal is not a spatially random field.  The spectral index
measured from the map of Galactic radio emission changes significantly
as the brightest regions of the Galactic plane are gradually masked
off. This does not happen when one applies the same type of
masking to a random field. We are planning to study the
level of Gaussianity of the Galactic signal from the Rhodes/HartRAO
survey data in a future work.

As it is apparent from Fig.~\ref{fig:filt2326}, the Rhodes/HartRAO
survey is affected by a certain level of striping, which reflects
baseline instabilities due to ground spillover and weather effects
(\cite{Jonas98}). This is more visible around the South Pole region,
where the scans necessarily traversed low elevation angles.  This
striping, with a typical angular scale of $\sim 5^{\circ}$, will
introduce some additional power in the spectrum at multipole order $l
\sim 40$. However, as apparent from the map, even at high-Galactic
latitudes, the amplitude of these striations are on the average lower
than the observed variation of sky brightness temperature, therefore we
expect striations to be only a minor contribution to the overall power
measured around $l \sim 40$.  The fact that no significant change of
slope is observed in the APS after point-source suppression for $l <
100$ (Fig.~\ref{fig:compFilt}), is another indication that striping
makes a minor contribution to the total power at a given angular scale.
If striping does make a major contribution, then its effect is being
disguised by a steeply falling power from the Galactic diffuse emission.
So, for the purpose of estimating the contribution of synchrotron
emission relatively to the CMB anisotropy, a spectral index of $\alpha =
2.9$ for the synchrotron APS is a conservative estimate.

\begin{figure}[thbp]
  \begin{center}
    \leavevmode \epsfig{file=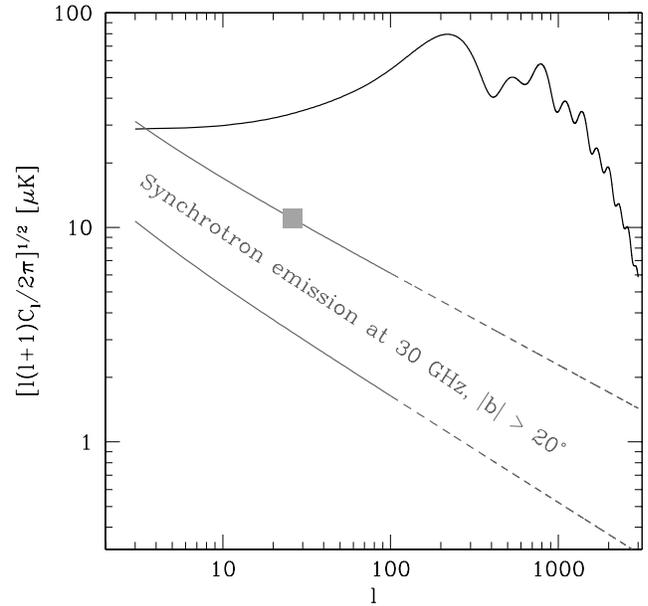, width=9.0cm, clip=}
    \caption{The estimated range for the APS of synchrotron emission at
      30 GHz (grey line) can be compared here with the APS of the CMB
      (black line), as derived with the latest estimate for the
      cosmological parameters (see text). In the $l$-range where it has
      been extrapolated the APS of synchrotron emission is shown as a
      dashed line. The square point is the upper limit to synchrotron
      temperature fluctuations derived from COBE DMR at 31.5, for the
      $7^{\circ}$ angular scale.}
    \label{fig:Extr30GHz}
  \end{center}
\end{figure}

One can use the APS derived at 2.3 GHz to infer the level of
contamination of the CMB signal at the different angular scales at 30
GHz: this is the lowest frequency channel of Planck and the one
for which the highest level of synchrotron contaminations are expected.
On the $7^{\circ}$ scale, \cite*{Kogut96b} derived an upper limit of 11
$\mu K$ to the temperature fluctuations due to synchrotron emission in
the COBE DMR channel at 31.5 GHz for Galactic latitudes with $|b| >
20^{\circ}$.  If we combine this limit with our measure of $\alpha =
2.92 \pm 0.07$ for the spectral index of the APS we derive an upper
limit of $A_{\rm 30 GHz} = 120~\mu$K for the amplitude of synchrotron
emission .  This upper-limit to synchrotron emission is shown against
a CMB spectrum in Fig.~\ref{fig:Extr30GHz}, as a function of the
multipole order $l$.  The APS for the CMB has been computed assuming a
flat inflationary model with purely scalar scale-invariant
fluctuations, vacuum energy $\Omega_{\Lambda} = 0.43$, cold dark matter
density $h^2\Omega_{\rm cdm}=0.20$ and baryon density $h^2\Omega_{\rm
  b}=0.03$, (where $h = 0.63$ is the Hubble parameter), as derived by
\cite*{Tegmark00b} from the latest CMB measurements
(\cite{DeBernardis00}; \cite{Winant00}). From Fig.~\ref{fig:Extr30GHz},
it is apparent that synchrotron emission can be a significant
contribution to the sky anisotropy at 30 GHz at scales larger than
$\sim 20^{\circ}$, but it is an order of magnitude weaker than the
cosmological signal at degree angular scales where the first
``acoustic'' peak of the CMB is observed.
Indeed from COBE observations we know that for the quadrupole (l=2),
Galactic emission is comparable in amplitude to the anisotropy of the
CMB (\cite{Bennett96}).

If $-\beta$ is the spectral index of the {\em frequency} dependency of
$A$, then $A_{\rm 30 GHz} = 120~\mu$K combined with the value $A_{\rm
  2.3 GHz} = 0.3\pm 0.2$~K measured at 2.3 GHz provides a lower limit
on the frequency spectral index of $\beta > 2.9$, between 2.3 and 31.5
GHz.  At large angular scales ($18^{\circ}$), \cite*{Platania98} derive
an average value for the spectral index of the synchrotron frequency
dependency between 1 and 7.5 GHz of $\beta = 2.81\pm 0.16$.  The
spectral index is expected to steepen at frequencies higher than 10
GHz, since the observed energy spectrum of cosmic ray electrons
steepens at energy of $\sim 10-20$ GeV (\cite{Banday90};
\cite{Banday91}).

In order to bracket the synchrotron contribution to the sky
fluctuations at 30 GHz one can combine the lowest ($1\sigma$) boundary
of our measure for the amplitude, $A_{\rm 2.3 GHz} = 0.1$~K, with the
lowest boundary of the frequency spectral index derived by
\cite*{Platania98}, that is $\beta = 3$: one obtains $A_{\rm 30 GHz} =
40~\mu$K. This lower limit on the synchrotron contribution is also shown
in Fig.~\ref{fig:Extr30GHz} as a function of the multipole order.

\cite*{Tegmark00} give three possible estimates of $A$ for Galactic
synchrotron emission at 19 GHz and label them as optimistic,
pessimistic and ``middle-of-road''. Their pessimistic value of 192
$\mu$K at 19 GHz with a frequency spectral index of $\beta = 2.6$,
corresponds to a $A_{\rm 30 GHz} \sim 60~\mu$K which is within the
range of our estimate, but somewhat on the lower bound.

\section{Summary and Conclusions}
\label{sec:concl}

In order to analyse the global angular power spectrum of Galactic radio
emission we have re-sampled the Rhodes/HartRAO Galactic survey at 2326
MHz into a HEALPix map with {\em nside}$=512$. The point sources were
removed from the map by median filtering.  We find that a best-fit to
the angular power spectrum of the entire Rhodes/HartRAO survey with a
power-law spectrum $C_l \propto l^{-\alpha}$, gives $\alpha = 2.43 \pm
0.05$ in the $l$ range $1-100$. At high Galactic latitudes, $|b| >
20^{\circ}$, where CMB observations are carried out, the APS of the
radio emission steepens and it is well described by a power-law with
spectral index $\alpha = 2.92 \pm 0.07$ and amplitude $A_{\rm 2.3 GHz} =
0.3\pm0.2$~K, up to $l=100$. At these latitudes radio emission is
dominated by synchrotron emission. By combining our results with the
upper limit of 11~$\mu$K on the synchrotron contribution to temperature
fluctuations at 31.5 GHz, obtained by COBE-DMR for the 7-degree
angular scale, we derive an upper limit of $A_{\rm 30 GHz} = 120~\mu$K
to the amplitude of synchrotron emission at high Galactic latidue. This
in turn implies a lower limit of $\beta > 2.9$ on the spectral index of the
{\em frequency} dependency of $A$.

With an APS index of $\alpha = 2.9$ and an amplitude $A_{\rm 30 GHz}
< 120~\mu$K, at 30 GHz synchrotron emission could be a contaminant of
CMB observations at large angular scales ($l\leq 10$), but will not
obstruct a clean view of the CMB anisotropy at degree angular scales.

A map of point sources from the Rhodes/HartRAO survey was also
obtained. The APS at high Galactic latitude was compared with the
prediction derived from extra-galactic source counts, as interpolated
between 1.4 and 5 GHz, under the assumption of a purely random sky
distribution. The same source counts and the randomness assumption are
the bases of the studies that predict the level of temperature
fluctuation to be expected from extra-galactic point sources in CMB
observations (\cite*{ftdz}; \cite*{Toffolatti98}). We conclude that at
2.3 GHz and for a resolution of 20 arcmin, predictions of the point
source contribution and observations are consistent with each other.

The simulations that we have performed to evaluate the accuracy of the
results show that the spectra derived from the Rhodes/HartRAO survey
in the $l$-range 2--100 are reliable. Incomplete sky coverage does not
compromise the accuracy with which one can derive the APS of the field
by spherical harmonic decompoistion, if this field has a spectrum with
spectral index $\alpha  <3$. However we have also found that the
quality of the spectrum derived degrades significantly for spectral
index $\alpha \ga 3$, if large areas of the sky are missing.

The simulations showed that median filtering can be used to effectively
remove point sources and derive the APS of the diffuse component up to
$l=100$, for a beam FWHM of 20 arcmin.

\begin{acknowledgements}
 
  We thank U. Seljak \& M. Zaldarriaga for their CMBFAST software,
  which was used to generate the CMB angular power spectrum.

\end{acknowledgements}

\end{document}